\documentclass[reprint,aps,prl,showkeys,groupedaddress,longbibliography,floatfix]{revtex4-1}

\usepackage{graphicx}        
\usepackage{grffile}         
\usepackage{graphics}
\usepackage{epsfig}
\usepackage{verbatim}
\usepackage{dcolumn}        
\usepackage{bm}            
\usepackage{amsmath}
\usepackage{amssymb}
\usepackage{mathtools}
\usepackage{physics}
\usepackage{hyperref}  
\hypersetup{colorlinks,breaklinks,
            linkcolor=black,urlcolor=black,
            anchorcolor=black,citecolor=black}
\usepackage{xspace}
\usepackage{float}
\newcommand{\EQ}[1]{Eq.~(\ref{eq:#1})}
\newcommand{\EQS}[2]{Eqs.~(\ref{eq:#1}) and (\ref{eq:#2})}
\newcommand{\FIG}[1]{Fig.~\ref{fig:#1}}

\def\bal#1\eal{\begin{align}#1\end{align}}
\usepackage{color}
\usepackage[dvipsnames]{xcolor}

\usepackage[english]{babel}
\begin{document}
\title{Mpemba effect in terms of mean first passage time}

\author{Matthew R. Walker}
\affiliation{Department of Physics, University of Virginia, Charlottesville, VA 22904, USA}
\author{Marija Vucelja}
\email{mvucelja@virginia.edu}
\affiliation{Department of Physics, University of Virginia, Charlottesville, VA 22904, USA}
\affiliation{Department of Mathematics, University of Virginia, Charlottesville, VA 22904, USA}

\begin{abstract}
The Mpemba effect occurs when a system prepared at a hot temperature cools down faster to the bath temperature than an identical system starting at a warm temperature. We derive the condition for the Mpemba effect in the small-diffusion limit of overdamped Langevin dynamics on a double-well potential. Our results show the strong Mpemba effect occurs when the probability of being in a well at initial and bath temperature match, which agrees with experiments. We also derive the conditions for the weak Mpemba effect and express the conditions for the effects in terms of mean first passage time.
\end{abstract}


\maketitle
Rapid cooling or heating of a physical system can lead to unusual thermal relaxation phenomena. A prime example of anomalous thermal relaxation is the Mpemba effect. The phenomenon occurs when a system prepared at a hot temperature overtakes an identical system prepared at a warm temperature and equilibrates faster to the cold environment~\cite{lu_nonequilibrium_2017}. A related effect exists in heating~\cite{lu_nonequilibrium_2017,kumar_anomalous_2022}. Comparing two identical physical systems in their relaxation to the environment, one expects that the system with a smaller mismatch between its own and the environment's temperature would thermalize faster -- yet it is not always the case. The Mpemba effect has been observed: water~\cite{jeng_mpemba_2006,mpemba_cool_1969}, clathrate hydrates~\cite{ahn_experimental_2016}, magnetic systems~\cite{chaddah_overtaking_2010}, polymers~\cite{hu_conformation_2018}, and colloidal particle systems~\cite{kumar_exponentially_2020}. Numerically it has been seen in spin-glasses~\cite{baity-jesi_mpemba_2019}, systems without equipartition~\cite{gijon_paths_2019}, driven granular gasses~\cite{gomez_gonzalez_time-dependent_2021,mompo_memory_2021,lasanta_when_2017,megias_mpemba-like_2022,biswas_mpemba_2022,biswas_mpemba_2022-1,torrente_large_2019,biswas_mpemba_2020}, cold gasses~\cite{keller_quenches_2018}, quantum systems~\cite{nava_lindblad_2019,kochsiek_accelerating_2022,carollo_exponentially_2021}, and antiferromagnets~\cite{lu_nonequilibrium_2017, klich_mpemba_2019, teza_relaxation_2021,teza_far_2022,teza_eigenvalue_2023}. The copious observations imply that the effect is general. It was studied in several theoretical works~\cite{lu_nonequilibrium_2017,klich_mpemba_2019,walker_anomalous_2021,busiello_inducing_2021,degunther_anomalous_2022,lin_power_2022,holtzman_landau_2022}. 

The prevalence of the effect suggests that in understanding such "shortcuts" to thermalization, we might gain insight into a general aspect of nonequilibrium statistical mechanics. On the practical level, the Mpemba effect is closely tied to optimal heating and cooling protocols and efficient sampling of phase spaces. It is thus of broad interest to industry and science to characterize this effect. To study the Mpemba effect as a paradigm, we use the overdamped Langevin dynamics on a double-well potential. Our enabling example has wide applications, from chemical reactions, polymers, colloids, escapes of metastable states, models of quantum tunneling, and scalar field theories. For a classical point particle in a potential with a metastable state this problem is well studied Kramers' escape problem~\cite{kramers_brownian_1940,melnikov_kramers_1991,langer_statistical_1969,moss_noise_1988,pontryagin_statistical_1933}. However, even in this simple setting, it is unknown when the Mpemba effect occurs. 

This letter derives the necessary conditions for the Mpemba effect in the small-diffusion limit of overdamped Langevin dynamics on a double-well potential. The condition for the effect is expressed in terms of mean first passage times. Our results agree with the experiments of Kumar and Bechhoefer~\cite{kumar_exponentially_2020}, who looked at the Mpemba effect for a thermal quench of a colloidal system in water. 

The plan of the paper is as follows. First, we introduce the model and the Mpemba effect. Next, we derive the necessary conditions for the Mpemba effect. Our analytical results are in the small-diffusion and large barrier limit, where the Kramers' problem is analytically solvable.  

\emph{Model} -- We consider an overdamped Langevin dynamics of a particle on a double-well potential $U$, schematically shown on~\FIG{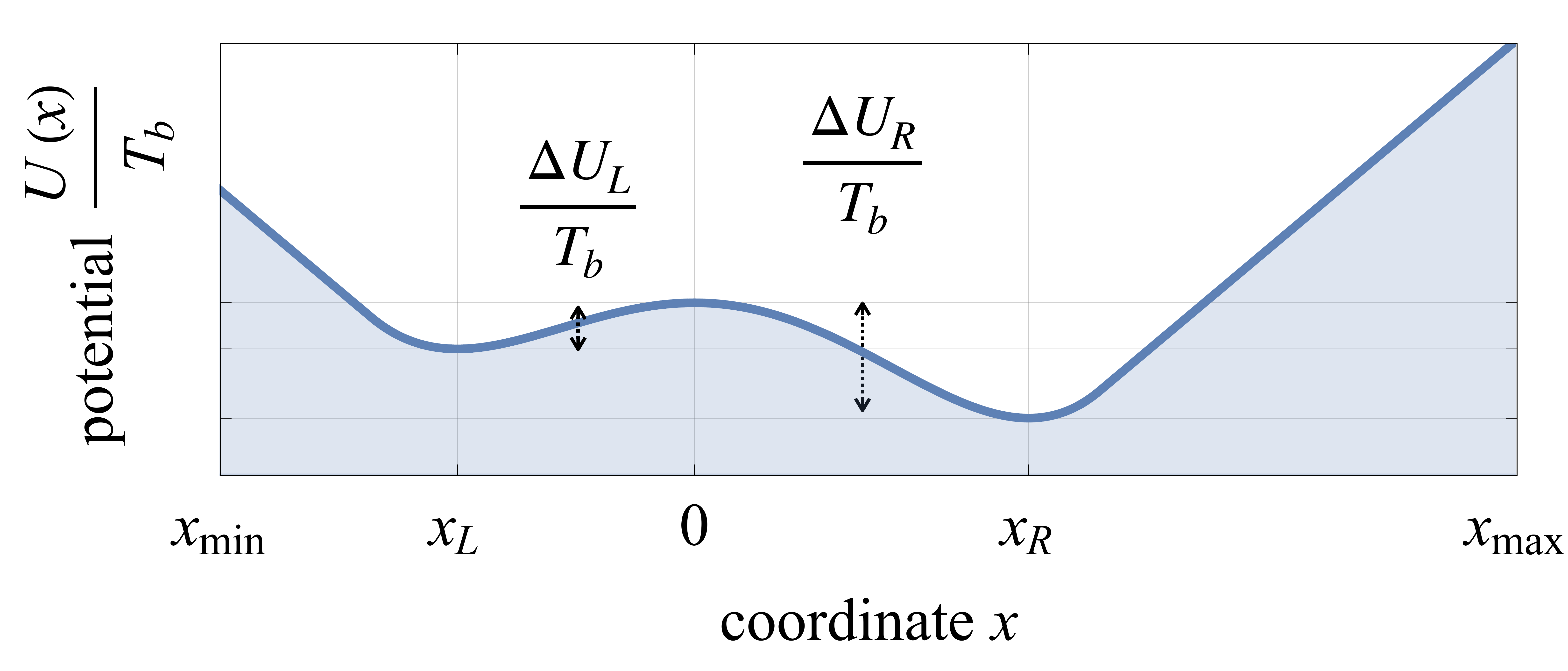}, and in a thermal bath of solvent molecules. The particle's trajectory ${\rm x}(t)$ obeys
\begin{eqnarray}
    \label{eq:Langevin-eq}
      \gamma \frac{d}{dt}{\rm x}(t) = - \frac{1}{m}U'[{\rm x}(t)] + \Gamma(t), 
\end{eqnarray}
where $-U' \equiv -dU/d{\rm x}$ is the force, $\gamma$ is the friction coefficient, and $\Gamma (t)$ is the thermal noise per unit mass. At times much larger than the collision time between the particle and the solvent $\Gamma (t)$ obeys Gaussian statistics, with $\mathbb{E}[ \Gamma(t) ]= 0$ and $\mathbb{E} [\Gamma (t) \Gamma (t')] = 2 \gamma (k_B T_b/m) \delta (t-t')$. The diffusion coefficient is $k_BT_b/m\gamma$. Below, we set the Boltzmann constant and particle mass to unity ($k_B = 1$, $m = 1$). The corresponding Fokker-Planck (FP) equation describes the evolution of a probability density, $p(x,t)$, of having a particle at time $t$ at coordinate $x$,
\begin{eqnarray}
    \label{eq:FPeq}
    \partial _t p(x,t) &= \frac{1}{\gamma}\partial _x \left[U'(x) + T_b\partial_x\right]p(x,t) \equiv \mathcal{L}p(x,t),
\end{eqnarray}
where $\mathcal{L}$ is the FP operator. We also denote the probability current density, $j(x,t)$,  as $\partial_t p(x,t) \equiv - \partial_x j(x,t)$. The system is closed and we thus have reflective boundary conditions. The stationary distribution 
is the Boltzmann distribution,   
\begin{eqnarray}
    \pi_{T_b}(x) =\frac{1}{Z(T_b)} e^{-\frac{U(x)}{T_b}},
\end{eqnarray}
where $Z(T_b) \equiv \int _{x_{\rm min}} ^{x_{\rm max}} \exp[-U(x)/T_b]\,dx$ is the partition function. 
\begin{figure}
    \centering  \includegraphics[width=\columnwidth]{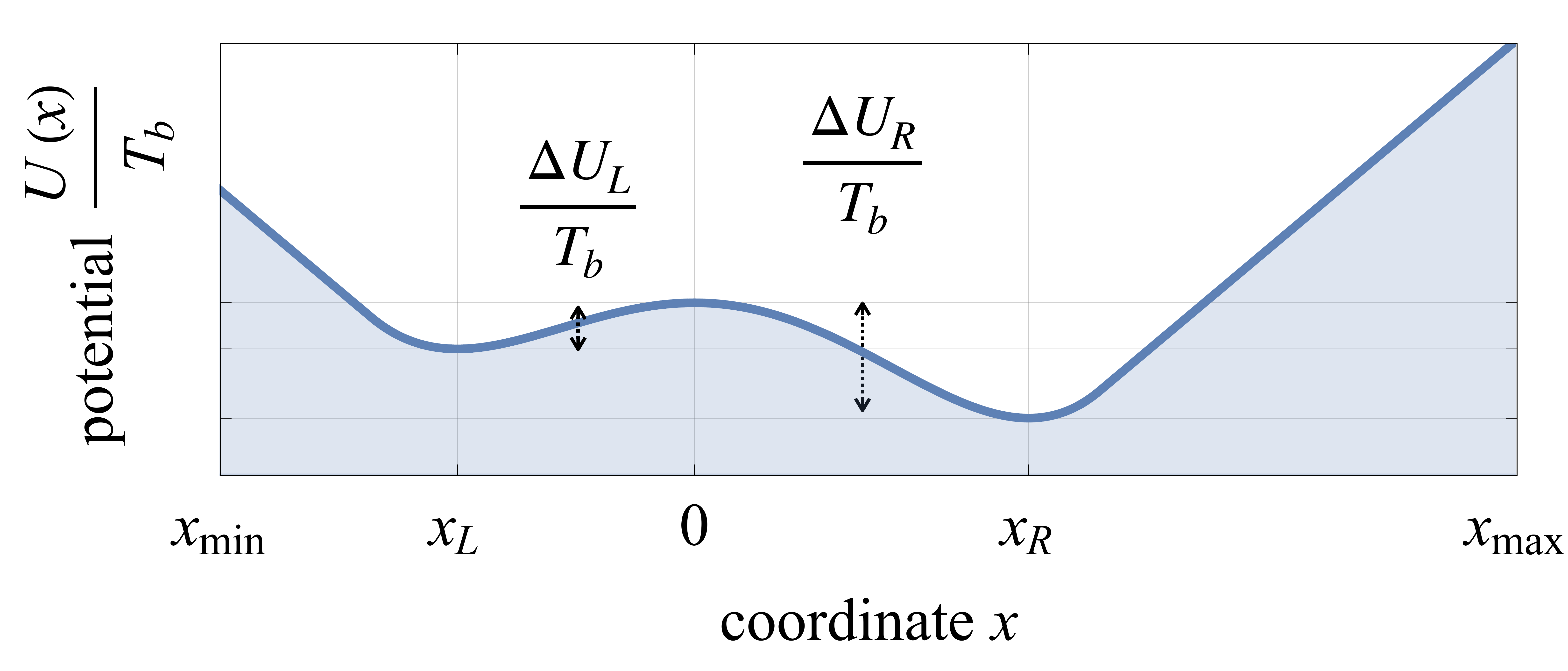}
    \caption{Double-well potential $U(x)$ with the barrier centered at $x = 0$, and minima at $x_L$ and $x_R$. The barrier heights are $\Delta U_L$ and $\Delta U_R$.}
    \label{fig:fig-potential-sketch.pdf}
\end{figure}

The scalar product is defined as $\langle u, v \rangle \equiv \int _{x_{\rm min}}  ^{x_{\rm max}}u(x)v(x)\,dx$ and the adjoint operator of the FP operator is $\mathcal{L}^\dagger = \gamma^{-1}[- U'\partial _x + T_b\partial^2 _x]$. The corresponding eigenvalue problems are: $\mathcal{L} v_i = \lambda_i v_i$ and $\mathcal{L}^\dagger u_i = \lambda_i u_i$. The two eigenfunctions are related as $u_i(x) = \exp[U(x)/T_b]v_i(x)$. The eigenvalues are ordered with $\lambda_1 = 0$ being the eigenvalue corresponding to $\pi_{T_b}$ and others are negative: $0 > \lambda_2 \geq \lambda_3 \geq \dots$ The probability density $p(x,t)$ is 
\begin{eqnarray}
\label{eq:pdf-expansion}
    p(x,t) = \pi_{T_b}(x) + \sum _{i>1}  a_i e^{\lambda_i t} v_i(x) , 
\end{eqnarray}
with overlap coefficients
\begin{eqnarray}
    \label{eq:ai-coeff}
    a_i \equiv \frac{\langle u_i, p_{\rm init} \rangle}{\langle u_i, v_i\rangle},  
\end{eqnarray}
where $p_{\rm init}$ is the initial condition. 

\emph{Strong and weak Mpemba effect} -- We assume that the system starts from equilibrium at temperature $T$, $\pi_T$, and consider cases with a gap between the second and the third eigenvalue, $\lambda_2 > \lambda_3$. The strong Mpemba effect happens when the overlap $a_2$, defined in~\EQ{ai-coeff}, is zero, i.e., when 
\begin{eqnarray}
    \label{eq:SMEcondu2pi}
	\langle u_2 \rangle _{T} = 0,
\end{eqnarray}
where $\langle \cdot \rangle_T$ is the equilibrium expectation value at temperature $T$. The strong Mpemba effect is characterized by a jump of relaxation time from $-\lambda_2 ^{-1}$ to $-\lambda_3 ^{-1}$, and hence exponentially faster relaxation toward equilibrium. It occurs for initial conditions that are orthogonal to the slowest relaxation mode. The strong Mpemba effect was introduced by Klich, Raz, Hirschberg, and Vucelja in~\cite{klich_mpemba_2019}, and experimentally first observed by Kumar and Bechhoefer~\cite{kumar_exponentially_2020}. The strong Mpemba effect implies the weak Mpemba effect, which happens when $a_2$ is a non-monotonic function of initial temperature $T$~\cite{lu_nonequilibrium_2017}, i.e., when $\partial_T a_2 = 0$, which for finite $T$ reduces to 
\begin{eqnarray}
\label{eq:WMEcondu2pi}
    \langle u_2 U \rangle _T - \langle u_2 \rangle _T \langle U \rangle _T = 0. 
\end{eqnarray}
Next we find $\lambda_2$ and $u_2$. 

\emph{Spectrum of the adjoint FP operator} -- For the spectrum of the adjoint FP operator, we look at the following eigenvalue problem 
\begin{eqnarray}
    \partial _x e^{-\frac{U(x)}{T_b}}\partial_x u_i(x) = \frac{\gamma\lambda_i}{T_b} e^{-\frac{U(x)}{T_b}}u_i(x).
\end{eqnarray}
Integrating the equation from $x_{\rm min}$ to $x$ twice and using the conservation of probability, we have 
\begin{eqnarray}
    \nonumber
    &u_i(x)= u_i(x_{\rm min})\times 
    \\
    \label{eq:uiL-eig-vector-v01}
   &\times
    \left[1+\frac{\gamma\lambda_i}{T_b}\frac{\int ^x _{x_{\rm min}}  e^{\frac{U(y)}{T_b}} \, dy \int ^y _{x_{\rm min}}  e^{-\frac{U(z)}{T_b}}u_i(z)\, dz }{ u_i(x_{\rm min})} \right].
\end{eqnarray}
For details, see the  supplementary material. Similarly integrating twice from $x$ to $x_{\rm max}$ and using  the boundary condition $u'_i(x_{\rm max}) = 0$, we get another expression for $u_i$, 
\begin{eqnarray}
    \nonumber
    &u_i(x)= u_i(x_{\rm min})\alpha_i\times 
    \\
     \label{eq:uiR-eig-vector-v01}
   &\times
   \left[1+\frac{\gamma\lambda_i}{T_b}\frac{\int ^{x_{\rm max}} _x  e^{\frac{U(y)}{T_b}}\, dy \int ^{x_{\rm max}} _y e^{-\frac{U(z)}{T_b}}u_i(z)\,dz  }{u_i(x_{\rm max})} \right],  
\end{eqnarray}
with $\alpha_i \equiv u_i(x_{\rm max})/u_i(x_{\rm min})$.     

In the small-diffusion limit the eigenfunction $u_i$ over $\mathcal{D}_L\equiv [x_{\rm min},0]$ is better approximated starting from~\EQ{uiL-eig-vector-v01} than~\EQ{uiR-eig-vector-v01}, and over $\mathcal{D}_R\equiv[0,x_{\rm max}]$ it is best to use \EQ{uiR-eig-vector-v01}. By demanding $u_i$ is continuous at $x = 0$ we get the eigenvalue 
\begin{eqnarray}
\nonumber
    \frac{\gamma \lambda_i}{T_b} =& (1-\alpha_i)\left[\alpha_i \frac{\int ^{x_{\rm max}} _0  e^{\frac{U(y)}{T_b}} \,dy\int ^{x_{\rm max}} _y  e^{-\frac{U(z)}{T_b}}u_i(z)\, dz}{u_i(x_{\rm max})} - \right.
    \\ 
    \label{eq:lambda-exact}
    &\left. - \frac{\int ^0 _{x_{\rm min}}  e^{\frac{U(y)}{T_b}}\,dy \int ^y _{x_{\rm min}}  e^{-\frac{U(z)}{T_b}}u_i(z)\,dz}{u_i(x_{\rm min})}\right]^{-1}. 
\end{eqnarray}
Below we derive the first nonzero eigenvalue, $\lambda_2$, and then the corresponding left eigenfunction, $u_2$. 

The magnitude $-\lambda_2$ signifies the switching rate between the two wells. The so-called Kramers problem is analytically solvable in the limit of small-diffusion and large barriers, c.f.~\cite{risken_fokker-planck_1996,zwanzig_nonequilibrium_2001}. Both barriers, shown on~\FIG{fig-potential-sketch.pdf}, should be large compared to diffusion, i.e., $|\Delta U_L| \gg T_b$ and $|\Delta U_R| \gg T_b$. In this limit, the transition rate between the wells is low, i.e., $\lambda_2$ is small. We use this to get $\lambda_2$ and $u_2$. In the zeroth approximation, diffusion is so small that there are no jumps between the wells, 
$\lambda^{(0)}_2 = 0$, this implies, that $u_2$, defined in \EQS{uiL-eig-vector-v01}{uiR-eig-vector-v01}, is step function
\begin{eqnarray}
\label{eq:u2-zeroth}
u_2^{(0)} = \begin{cases}
    u_2(x_{\rm min}), & x \in \mathcal{D}_L
    \\
    u_2 (x_{\rm max}), & x \in \mathcal{D}_R
\end{cases}.
\end{eqnarray}
In this case, the ergodicity is broken. For what follows, it is important to note that we centered the potential so it has a local maximum at $x =0$. The coefficient $\alpha_2 ^{(0)}$ we get from demanding $u_2'(x)$ is continuous at $x = 0$. In the zeroth-order approximation 
we have
$
    \alpha^{(0)} _2 = -\int ^0 _{x_{\rm min}} e^{-\frac{U(z)}{T_b}}\,dz/\int ^{x_{\rm max}} _0 e^{-\frac{U(z)}{T_b}}\,dz \equiv - \Pi_L(T_b)/\Pi_R(T_b)$, 
where we label with $\Pi_{L}(T_b)$ and $\Pi_R(T_b)$ the probabilities of the particle being in left and right well at temperature $T_b$. Plugging in $u_2^{(0)}$,~\EQ{u2-zeroth}, in the exact expression for $\lambda_2$,~\EQ{lambda-exact}, we get 
\begin{eqnarray}
    \frac{\gamma \lambda_2 ^{(1)}}{T_b} 
    =&-\frac{Z(T_b)}{\Pi _R(T_b) \mathcal{A}_L(0)+\Pi _L(T_b) \mathcal{A}_R(0)}, 
\end{eqnarray}
where we denoted 
\begin{eqnarray}
    \label{eq:AR}
    &\mathcal{A}_R(x) \equiv \int ^{x_{\rm max}} _x  e^{\frac{U(y)}{T_b}}\,dy \int ^{x_{\rm max}} _y  e^{-\frac{U(z)}{T_b}}\,dz , 
    \\
    \label{eq:AL}
    &
    \mathcal{A}_L(x) \equiv \int ^x _{x_{\rm min}}  e^{\frac{U(y)}{T_b}}\,dy \int ^y _{x_{\rm min}}  e^{-\frac{U(z)}{T_b}}\,dz. 
\end{eqnarray}
The eigenvector is 
\begin{eqnarray}
\label{eq:u2-first-corr}
    u^{(1)}_2(x) \propto \begin{cases}
    1 + \frac{\gamma \lambda^{(1)}_2}{T_b}\mathcal{A}_L(x), &x\in \mathcal{D}_L 
    \\[10pt]
   \alpha^{(1)} _2 + \alpha^{(0)} _2 \frac{\gamma \lambda^{(1)}_2}{T_b}\mathcal{A}_R(x), &x\in \mathcal{D}_R 
    \end{cases}.
\end{eqnarray}
From continuity of $u_2(x)$ at $x = 0$, we have $\alpha^{(1)} _2 = \alpha ^{(0)} _2$. 

Note that the results for $\lambda_2$ and $u_2$ could have been obtained by solving for the ground state and the lowest eigenfunction of the adjoint FP operator with the inverted potential $- U(x)$ and absorbing boundary conditions since there is an exact mapping between the two problems, see e.g.~\cite{risken_fokker-planck_1996}.

Below we remain with the first nonzero corrections to $\lambda_2$. For simplicity, we drop writing the superscripts above $u_2$, $\lambda_2$, and $\alpha_2$. Before proceeding further, it is instructive to express $\mathcal{A} _L(x)$ and $\mathcal{A}_R(x)$ in terms of mean first passage times. 

\emph{Mean First Passage Time} -- A typical Mean First Passage Time (MFPT) scenario tracks particles that leave a domain for the first time and do not return to it, see e.g.~\cite{zwanzig_nonequilibrium_2001}. Suppose we look at the domain $\mathcal{D}_R$. The motion of particles is governed by the Langevin equation~\EQ{Langevin-eq}. Also, suppose our initial point $x_0 \in  \mathcal{D}_R$. The first passage time is the time the particle leaves the domain. To find the MFPT, we focus on the trajectories that have not left the domain $\mathcal{D}_R$ before time $t$. The distribution of such particles obeys the FP equation
\begin{eqnarray}
    \partial_t \tilde{p} = \tilde{\mathcal{L}} \tilde{p},  
\end{eqnarray}
where $\tilde{\mathcal{L}} = \mathcal{L}$, and we added the tilde symbol to signify different initial and boundary conditions from the rest of the paper. Here the initial condition is $\tilde{p}(x, 0) = \delta (x - x_0)$ for $x_0 \in \mathcal{D}_R$ and the boundary conditions are: $\tilde{j}(x_{\rm max}) = \left[\tilde{p}' + U' \tilde{p}\right]_{x = x_{\rm max}} = 0$ and $\tilde{p}(0,t) = 0$. 
The number of points that are still in $\mathcal{D}_R$ at time $t$ is 
\begin{eqnarray}
    \tilde{P}(t, x_0) = \int _{\mathcal{D}_R} \tilde{p}(x,t)\,dx. 
\end{eqnarray}
The number of points that have not left before time $t$ but have left during time integral $(t, t+ dt)$ is  
\begin{eqnarray}
    \tilde{P}(t, x_0) - \tilde{P}(t + dt, x_0) = \rho(t,x_0)dt, 
\end{eqnarray}
where $\rho(t,x_0)$ is the distribution of first passage times. MFPT is the first moment of $t$ of $\rho(t,x_0)$, i.e., 
\begin{eqnarray}
    \tau _R (x_0) = \int _0 ^{\infty} t \rho (t, x_0) \,dt=\int _0 ^{\infty} \tilde{P} (t, x_0)\,dt. 
\end{eqnarray}
In the 1D case, $\tau_R$ can be calculated explicitly (see the supplementary material) and leads to 
\begin{eqnarray}
\label{eq:DRMFPTeq}
    T_b \gamma^{-1}e^{\frac{U(x)}{T_b}}\partial_x e^{-\frac{U(x)}{T_b}}\partial_x \tau_R(x) = 
    -1, 
\end{eqnarray}
with boundary condition $\tau_R(0) = 0$, which means that any initial point on the boundary will leave immediately. Next, we assume that MFPT at $x = x_{\rm max}$ approaches a constant, i.e., $\tau'(x_{\rm max}) = 0$. Integrating~\EQ{DRMFPTeq}, twice for the right domain, we get 
\begin{eqnarray}
\label{eq:ARmeaning}
    \mathcal{A}_R(x) = T_b\gamma^{-1}\left[\tau_R(x_{\rm max}) - \tau_R(x)\right], 
\end{eqnarray}
for  $x \in \mathcal{D}_R$. 
The derivation for the left domain, $\mathcal{D}_L$, is analogous. The MFPT $\tau_L$ satisfies 
$   T_b\gamma ^{-1}e^{\frac{U(x)}{T_b}}\partial_x e^{-\frac{U(x)}{T_b}}\partial_x \tau_L(x) = 
    -1,$
with boundary conditions $\tau_L'(x_{\rm min}) = 0$ and $\tau_L(0) = 0$. We integrate the equation twice and for $x \in \mathcal{D}_L$ get
\begin{eqnarray}
\label{eq:ALmeaning}
    \mathcal{A}_L(x) = T_b \gamma^{-1}\left[\tau_L (x_{\rm min}) - \tau_L(x)\right]. 
\end{eqnarray}

Now, using~\EQS{ALmeaning}{ARmeaning} we can write the eigenvalue as 
\begin{eqnarray}
    \lambda_2 = - \frac{Z(T_b)}{ \Pi_R(T_b)\tau_L(x_{\rm min})+\Pi_L(T_b)\tau_R(x_{\rm max})}. 
\end{eqnarray} 
In the small-diffusion limit the exponential integrals in $Z(T_b)$, $\tau_R(x_{\rm max})$ and $\tau_L(x_{\rm min})$ are readily approximated by Laplace's method. In this limit, $\lambda_2$ reduces to the sum of Kramers' rates from one well to another (see supplementary material and c.f.~\cite{risken_fokker-planck_1996}). 

Now that we have obtained the expression for $u_2$ (see ~\EQ{u2-first-corr}) 
in terms of the MFPTs, 
we 
can write down the conditions for the Mpemba effect. 

\begin{figure}
    \includegraphics[width =\columnwidth]{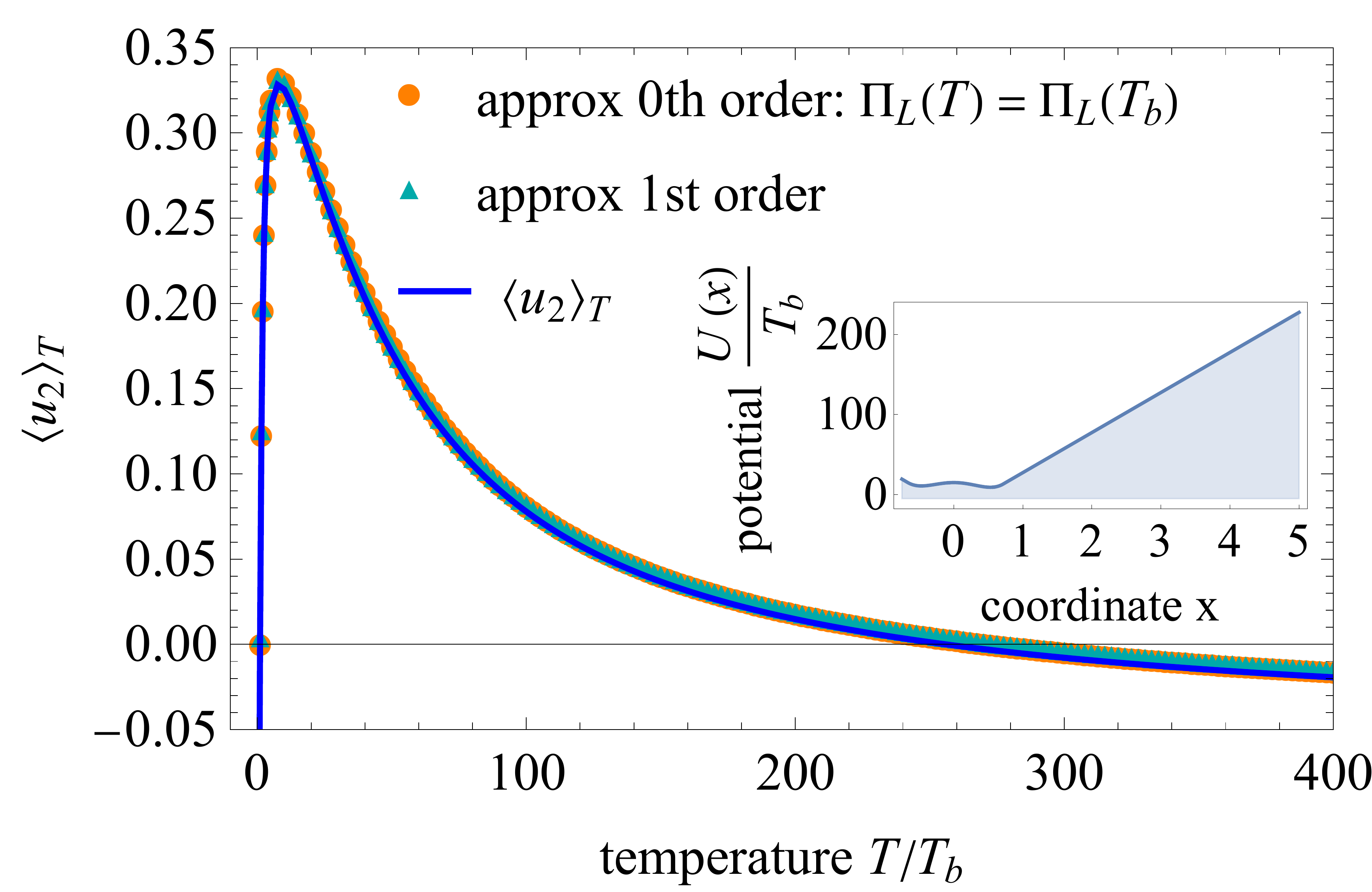}
    \caption{
    An example of a strong Mpemba effect at $T \approx 256 \, T_b$. The small-diffusion parameters are: $T_b/|\Delta U_L| = 0.24$ and $T_b/|\Delta U_R| = 0.17$. The eigenvalue is $\lambda_2 \approx - 0.287$ numerically and $\lambda^{(1)}_2 \approx - 0.284$ with our approximation. The inset shows the potential. The condition for the strong Mpemba effect stated in~\EQ{SMEcondu2pi} (blue line) and approximated with~\EQ{SMEcond} (orange circles are $0^{\rm th}$ order and cyan triangles are $1^{\rm st}$ order approximation). In the zeroth order, we neglected terms proportional to $\gamma \lambda_2/T_b$ in~\EQ{SMEcond}. Both orders agree well with the numerics.}
    \label{fig:fig-a2SME-v01.pdf}
\end{figure}

\emph{Condition for the strong Mpemba effect} -- Plugging in the expression for $u_2$, \EQ{u2-first-corr}, in~\EQ{SMEcondu2pi}, we get condition for the strong Mpemba effect
\begin{eqnarray} 
\nonumber
    &0=
    \left(\frac{\Pi _L (T)}{\Pi _L(T_b)} - \frac{\Pi _R (T)}{\Pi _R(T_b)}\right)  + 
    \\
& \label{eq:SMEcond}
+\frac{\gamma \lambda_2}{T_b}\left(\langle \mathcal{A}_L\rangle _{L,T}\frac{\Pi _L(T)}{\Pi _L(T_b)} - \frac{\Pi _R(T)}{\Pi _R(T_b)}\langle \mathcal{A}_R\rangle _{R,T}  \right)
, 
\end{eqnarray}
where $\langle \cdot \rangle_{X,T}$ is the average over $\mathcal{D}_X$ with probability distribution $\pi(T)Z(T)/\Pi_X(T)$, where $X$ is $L$ or $R$. For vanishing small $\gamma \lambda_2 /T_b$, and using $\Pi _L + \Pi _R = 1$, the above expression reduces to 
\begin{eqnarray}
\label{eq:SMEcondzeroth}
    \Pi _L (T)=\Pi _L (T_b),
\end{eqnarray}
which is what Kumar and Bechhoefer saw in experiment~\cite{kumar_exponentially_2020}. That is, they observed that the strong Mpemba effect occurs when the probability of the particle being in a well is the same for the initial and the bath temperature. Hence for vanishingly small $\gamma \lambda_2 /T_b$, the strong Mpemba effect occurs when all of the equilibrium probability of being in a well is already there at the beginning. A plausible rationale is that in the limit of large barriers, it is "faster" to "drape" the probability density inside the well differently than switch between wells. Thus the strong Mpemba effect occurs when we start from close to the "right amount" of probability in each well. Corrections linear in $\gamma \lambda_2/T_b$ and $\lambda_2$ give the dependence of the condition on MFPTs. An example of the strong Mpemba effect and use of~\EQ{SMEcond} is on \FIG{fig-a2SME-v01.pdf}.

\emph{Condition for the weak Mpemba effect} -- After plugging in $u_2$,~\EQ{u2-first-corr}, in~\EQ{WMEcondu2pi}, the necessary condition for the weak Mpemba effect is
\begin{eqnarray}  
    \label{eq:WMEcond}
    0=&W^{(0)} +\frac{\gamma \lambda_2}{T_b} W^{(1)}, 
\end{eqnarray}
where
\begin{eqnarray}
\nonumber
    W^{(0)} \equiv &\langle U\rangle _{L,T}\frac{\Pi_L(T)}{\Pi _L(T_b)} - \frac{\Pi _R (T)}{\Pi _R(T_b)} \langle U\rangle _{R,T} 
    \\
    &-\langle U\rangle _T \left(\frac{\Pi _L (T)}{\Pi _L(T_b)} - \frac{\Pi _R (T)}{\Pi _R(T_b)}\right),
\end{eqnarray}
and
\begin{eqnarray}
    \nonumber
    &W^{(1)} \equiv\langle U\mathcal{A}_L\rangle _{L,T}\frac{\Pi _L(T)}{\Pi _L(T_b)} - \frac{\Pi _R(T)}{\Pi _R(T_b)}\langle U\mathcal{A}_R\rangle _{R,T} 
    \\ \label{eq:WMEcondPitau1st}
    &-\langle U\rangle _T\left(\langle \mathcal{A}_L\rangle _{L,T}\frac{\Pi _L(T)}{\Pi _L(T_b)} - \frac{\Pi _R(T)}{\Pi _R(T_b)}\langle \mathcal{A}_R\rangle _{R,T}  \right).
\end{eqnarray}
For vanishingly small  $\gamma \lambda_2/T_b$ the condition for the Mpemba effect,~\EQ{WMEcond}, is $W^{(0)} = 0$. The dependence on MFPTs is in $\lambda_2$ and $W^{(1)}$. An example of the weak Mpemba effect and use of~\EQ{WMEcond} is on~\FIG{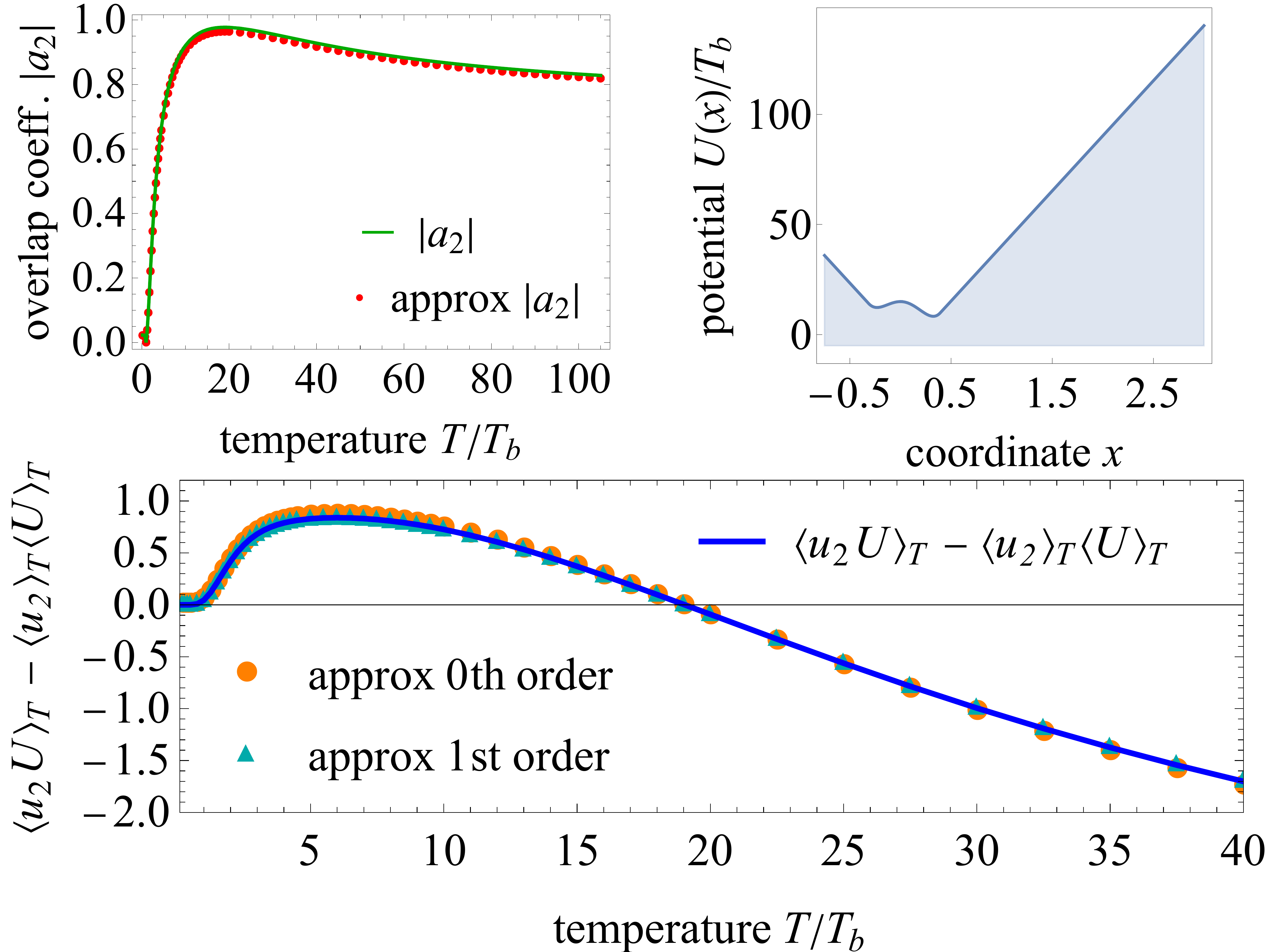}.
\begin{figure}
    \centering
    \includegraphics[width =\columnwidth]{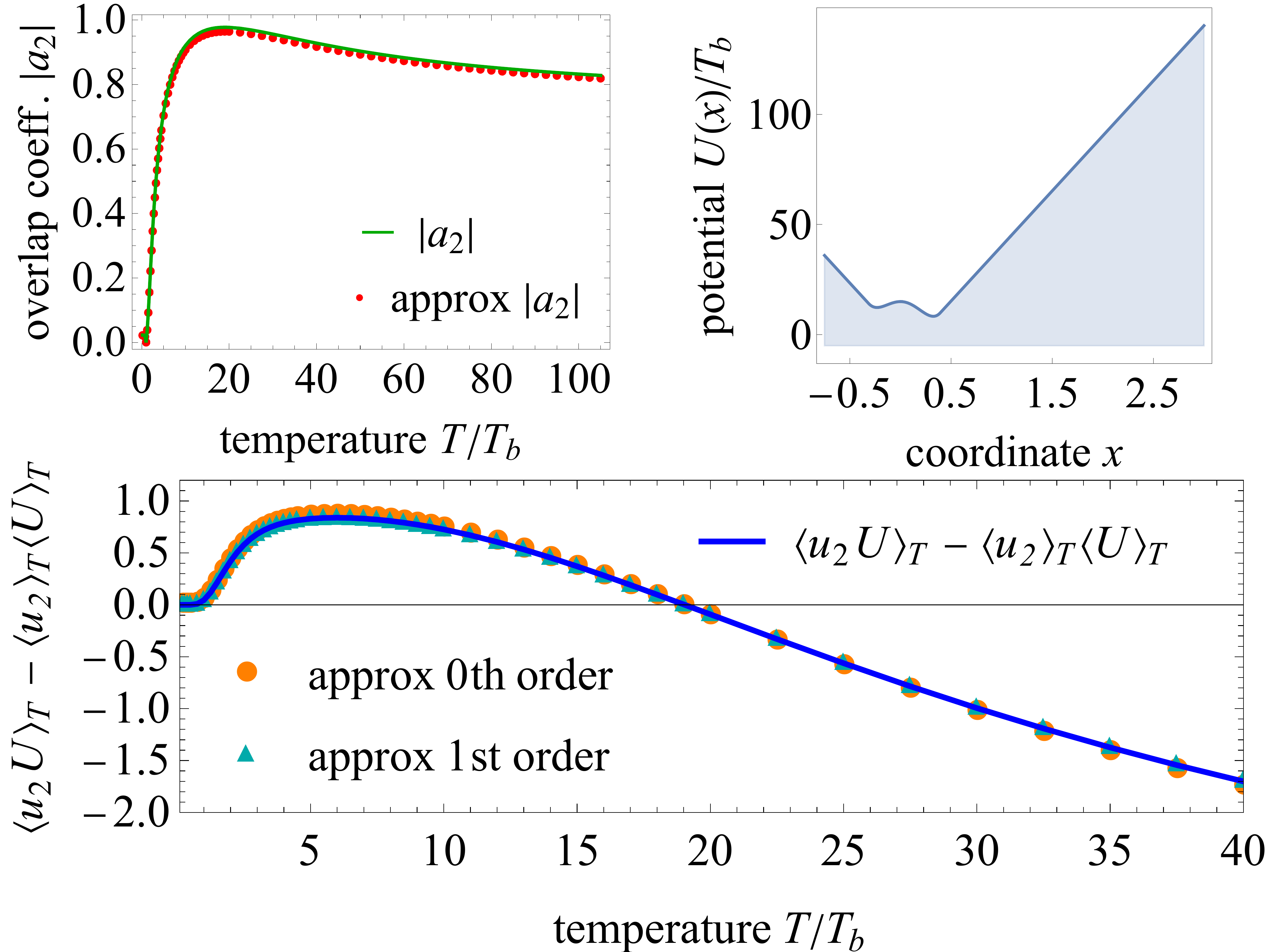}
    \caption{(top left) The overlap coefficient $a_2$ obtained numerically (green), and by using approximate  $u_2$,~\EQ{u2-first-corr} (red circles). At initial temperature $T \approx 19 \, T_b$ the overlap $a_2$ has a local maximum, which is the hallmark of the weak Mpemba effect. (top right) The potential with small-diffusion parameters: $T_b/|\Delta U_L| = 0.36$ and $T_b/|\Delta U_R| = 0.13$.  The eigenvalue is $\lambda_2 \approx - 0.92$ numerically and $\lambda^{(1)}_2 \approx - 0.89$ with our approximation. (bottom) The condition for the weak Mpemba effect stated in~\EQ{WMEcondu2pi} (blue line) and approximated with \EQ{WMEcond} (to $0^{\rm th}$ order: orange circles and $1^{\rm st}$ order: teal triangles). In the zeroth order, we neglected terms proportional to $\gamma \lambda_2/T_b$ in~\EQ{WMEcond}. Both orders agree well with the numerics.}
    \label{fig:fig-WMEapprox-v01.pdf}
\end{figure}

The necessary conditions for the Mpemba effect,~\EQS{SMEcond}{WMEcond}, express the relation between the MFPTs, mean energy, and the pair correlation of the MFPT and energy that must hold if the effect is to occur. These equations are the main result of this letter. 

\emph{No Mpemba effect for a two-level system} -- Notice that by reducing the problem of diffusion in a double-well potential to a two-level system with two states corresponding to the minima of the potential, we lose the Mpemba effect. In the case of a two-level system, one can see that~\EQS{SMEcond}{WMEcond} hold only if $T=T_b$. This result is expected as for the Mpemba effect; we need at least three eigenvectors and a gap ($\lambda_2>\lambda_3$). 

\emph{Generalizations} -- Our results generalize to the case of spatially-dependent diffusion and predict the Mpemba effect for potentials that are a multiple of the original potential. The approximate solution for the largest nonzero eigenvalue and eigenfunction can be generalized for multiple barriers. These generalizations are described below.  

\emph{Spatially dependent diffusion} -- For a 1D FP equation, the coordinate dependent diffusion coefficient can always be transformed to a constant $D(T_b) > 0$, see e.g.~\cite{risken_fokker-planck_1996}. If in the original coordinates, $\tilde{x}$, the diffusion coefficient was $\tilde{D}(\tilde{x},T_b)$, then in the new coordinates, $x$, it is $D(T_b) = \left(d x/d \tilde{x}\right)^2 \tilde{D}(\tilde{x},T_b)$. The transformation is $x(\tilde{x}) = \int ^{\tilde{x}}  _{\tilde{x}_0} d\tilde{y} (D(T_b)/\tilde{D}(T_b, \tilde{y}))^{1/2}$, where choice of $\tilde{x}_0$ determines the value of $D$. The transformed potential is 
\begin{eqnarray}
    - \frac{1}{\gamma} U'(x) = \frac{d x}{d \tilde{x}} \left(- \frac{1}{\gamma} \tilde{U}'(\tilde{x}) \right) + \left(\frac{d^2x}{d \tilde{x}^2}\right)
    \tilde{D}(\tilde{x},T_b). 
\end{eqnarray}
Due to this transform by knowing that there is a strong Mpemba effect ($a_2 = 0$) at $\{T_{\rm init} = T, T_b\}$, force $-U'$, and diffusion coefficient $D(T_b)$, we also know that there is a strong Mpemba effect for force $-\tilde{U}'(\tilde{x})$, diffusion coefficient $\tilde{D}(\tilde{x},T_b)$ and the same temperatures, $\{T_{\rm init} = T,T_b\}$. 

\emph{Scaling argument} -- Since the temperature always appears in a ratio with the potential, we have the same expression for the Boltzmann distribution and the eigenfunction $u_2$ for $\{U, T\}$ and for $\{\kappa U,\kappa T\}$, where $\kappa$ is a constant.  Thus a strong Mpemba effect for potential $U$, diffusion coefficient $T_b/\gamma$, and temperatures $\{T_{\rm init} = T,T_b\}$ implies a strong Mpemba effect for potential $\kappa U$, diffusion $\kappa T_b/\gamma$ and temperatures $\{T_{\rm init} = \kappa T,\kappa T_b\}$.

\emph{Multiple barriers} -- Note that the approximate method for the second eigenvalue and eigenfunction for the double-well potential readily generalizes for a potential with several minima within the small-diffusion limit. Deriving the conditions for the Mpemba effect for such potentials requires separate consideration. 

\emph{Discussion} -- We derive the necessary conditions for the Mpemba effect in the case of overdamped Langevin dynamics on a double-well potential. Our results predict the initial temperatures that lead to the Mpemba effect via integral equations that contain probabilities to be in the two wells and mean first passage times. The exponential integrals in the conditions can be readily evaluated by Laplace's method. For the strong Mpemba effect, our findings, in the leading order, agree with the experiments of Kumar and Bechhoefer~\cite{kumar_exponentially_2020}, who observed the strong Mpemba effect when the probabilities of being in a well at the initial temperature and the bath temperature match. With large barriers, it is "faster" to rearrange the probability within the well than to switch between wells; thus, it is plausible that the initial conditions leading to the strong Mpemba effect would be the one with the "right amount" (the same amount as in equilibrium) of probability in the wells. We also derive the conditions for the weak Mpemba effect, which would be wonderful to see in an experiment. 

The overdamped Langevin dynamics is a phenomenological description of many systems; thus it is interesting to see the physical interpretation of the Mpemba effect conditions in specific cases.  

\begin{acknowledgments}
This material is based upon work supported by the National Science Foundation under Grant No.~DMR-1944539. MV and MRW acknowledge discussions with Gregory Falkovich, David Mukamel, Baruch Meerson, Oren Raz, Gianluca Teza, Roi Holzmann, Aaron Winn, Shlomi Reuveni, Eli Barkai, and Peter Arnold. This work would not have been possible had MV not been invited to coffee at a critical moment to discuss anything and everything except the work of this paper.
\end{acknowledgments}

\bibliographystyle{apsrev4-1}
\bibliography{mpemba-instanton-bib}
\end{document}